\documentclass[twocolumn,prb,showpacs,multicol,amsmath,amssymb]{revtex4}
\usepackage[dvips]{graphicx}

\newcommand{\be}{\begin{equation}}
\newcommand{\ee}{\end{equation}}

\newcommand{\bea}{\begin{eqnarray}}
\newcommand{\eea}{\end{eqnarray}}
\newcommand{\bd}{\begin{displaymath}}
\newcommand{\ed}{\end{displaymath}}
\newcommand{\ba}{\begin{array}}
\newcommand{\ea}{\end{array}}
\newcommand{\bi}{\begin{itemize}}
\newcommand{\ei}{\end{itemize}}
\newcommand{\bc}{\begin{center}}
\newcommand{\ec}{\end{center}}
\newcommand{\bfl}{\begin{flushleft}}
\newcommand{\efl}{\end{flushleft}}
\newcommand{\bfr}{\begin{flushright}}
\newcommand{\efr}{\end{flushright}}

\def\csch{\mathop{{\rm csch}}}

\def\6{\partial}

\def\no{\nonumber \\}
\def\bra{\langle}
\def\ket{\rangle}
\def\={\!\!\!&=&\!\!\!}
\def\+{\!\!\!&&\!\!\!+~}
\def\-{\!\!\!&&\!\!\!-~}

\begin{document}
\date{10 March   2007}

\title{Quadrupole Effect  on the Heat Conductivity  of Cold Glasses}

\author {Alireza Akbari}

\affiliation{Max Planck Institute for the Physics of Complex
Systems, 01187 Dresden, Germany\\
Institute for Advanced Studies in
Basic Sciences, P.O.Box 45195-1159, Zanjan, Iran}

\begin{abstract}
At very low temperatures, the tunneling theory for amorphous
solids predicts a thermal conductivity $\kappa\propto T^p$, with
$p = 2$.
 We have studied the
effect of the Nuclear Quadrupole moment on the thermal
conductivity of glasses at very low temperatures. We developed  a
theory that couples the tunneling motion to the nuclear
quadrupoles moment in order to evaluate the thermal conductivity.
Our result suggests a  cross over between two different regimes
at the temperature close to the  nuclear quadrupoles energy. Below
this temperature we have shown that the thermal conductivity is
larger than the standard tunneling result and therefore we have $p
< 2$. However, for temperatures higher than the  nuclear
quadrupoles energy, the result of standard tunneling model has
been found.
\end{abstract}


\maketitle


\section{Introduction}

Amorphous or glassy materials differ significantly from crystals,
especially in the low temperature range. Below $1 \rm{K}$, the
specific heat $C_v$ of dielectric glasses is much larger
than in crystalline materials. Moreover, the
thermal conductivity $\kappa$ is orders of magnitude lower than the
corresponding values found in their crystalline counterparts.
$C_v$ depends approximately linearly and $\kappa$ almost
quadratically on temperature\cite{Zel71}. The generally accepted
basis to describe the low temperature properties of glasses is the
phenomenological tunneling model\cite{Phi72, And72}.
 To explain these behaviors, it was considered
that atoms, or groups of atoms, are tunneling between two
equilibrium positions, the two minima of a double well potential
(DWP). The model is known as the two level system (TLS). In the
standard TLS model, these tunneling excitations are considered as
independent, and some specific assumptions are made regarding the
parameters that characterize them\cite{Esq98}.

The TLS can be excited from its ground state to the upper level
therefore contributed to the heat capacity. TLSs can also scatter
phonons and in this way decrease their mean free path and,
correspondingly, the heat conductance.

New interest in this problem was stimulated by several experimental
results\cite{Str98,Str00,Woh01,Nag04}. Until these experiments it was the
general believe that the dielectric properties of insulating
non-magnetic glasses are independent of external magnetic field.
It is very surprising that strong magnetic field effects were
discovered in polarization echo experiments at radio
frequency and in low frequency dielectric susceptibility
measurement at very low temperatures \cite{Str98,Str00,Woh01}.
Several generalizations of the standard TLS model have been
reported after the anomalous behavior of glasses in a magnetic
field. According to these solutions, the models can be divided
into "orbital"\cite{Ket99,Wue02,Lan02,Akb05} and "spin" models
(nuclear quadrupole effect)\cite{Wue02a,Wue04,Dor03,Akb06}. The
"orbital" models can provide an explanation for some of the
magnetic field effects by considering the flux dependence of the
tunneling splitting. Unfortunately, some assumptions have been
made which cannot be reconciled with the standard features of the
tunneling model.

A surprising outcome of these experiments is a novel isotope
effect observed in different glasses \cite{Nag04}. The latter
effect shows the important influence of the nuclear quadrupole
moments on the observed magnetic field dependence. Therefore it is
very important to find the effect of nuclear quadrupole moments on
the response function of glasses. For this purpose, in this paper
we have studied the thermal properties of heat conductivity  of
cold glasses taking into account the quadrupole effects. In
Section II  using W\"{u}rger's formalism\cite{Wue04}, we introduce
the nuclear spins in the frame of the two level system model. We
will find the general form of the heat conductivity of cold
glasses which takes into account the nuclear quadrupole moment
 in Section III. And finally in section IV, we end this paper by a summery and conclusion on our
results.

\section{TLS coupled by  a nuclear spin}

The standard TLS can be described as a particle or a small group
of particles moving in an effective double-well potential. At very
low temperatures only the ground states of each
wells are relevant. Using a pseudo-spin representation the
Hamiltonian of such a TLS read as
\begin{equation}
H_{TLS} \; = \; \frac{1}{2}\Delta _{0}\sigma _{x} +
\frac{1}{2}\Delta \sigma _{z},
\label{eq1}
\end{equation}
where $\Delta $ is the energy off-set at the bottom of the
wells, and  $\Delta _{0}$ is the tunnel matrix element.
Diagonalization of this two state Hamiltonian gives the energies
\begin{equation}\nonumber
E_{\pm} \; = \;\pm\frac{1}{2}E=\pm\frac{1}{2} \sqrt{\Delta
_{0}^{2}+ \Delta ^{2}}
\label{eq2}
\end{equation}
where $E$ is the energy difference between the two wells.
According to the randomness of the glassy structure, the energy
difference between the two wells have a broad distribution. The
energy off-set and the tunneling matrix element are widely distributed and are independent of
each other with a uniform distribution of
\begin{equation}
{\cal P}\left( \Delta ,\Delta _{0}\right) \; = \;
\frac{P_{0}}{\Delta _{0}}
\label{eq3}
\end{equation}
where $P_0$ is a constant. Using the notations
$ u  = \frac{\Delta_0}{E} ; w = \frac{\Delta}{E}$
which satisfy $u^2 + w^2 =1$,
the corresponding eigenstates of the diagonal Hamiltonain are
given by
\bea |\psi_{\pm}\ket=\sqrt{\frac{1\pm
w}{2}}|L\ket\pm\sqrt{\frac{1\mp w}{2}}|R\ket
\label{eq6}
\eea
where $|\psi_{-}\ket$ and $|\psi_{+}\ket$  are the ground and
exited state of the system, respectively. For the moment there is
no rigorous theory for tunneling in glasses. It is assumed that
atoms or groups of atoms participate in one TLS. As we mentioned
before, in the case of the multi-component glasses, one or several
of the tunneling atoms carry a nuclear magnetic dipole and an
electric quadrupole. When the system moves from one well to
another, the atoms change their positions by a fraction of an \AA
ngstr\"{o}m.

We can describe the internal motion of the nuclei by a nuclear
spin ${\bf I}$ of absolute value ${\bf I}^2=\hbar^2I(I+1)$. For a
nucleus with spin quantum number $I \geq 1$ the charge
distribution $\rho({\bf r})$ is not isotropic. Beside the charge
monopole, an electric quadrupole moment can be defined with
respect to an axis ${\bf e}$
\begin{equation}
Q=\int d^{3}r\left[ 3({\bf r\cdot e})^{2}-{\bf r}^{2}\right] \rho
({\bf r}).
\label{eq7}
\end{equation}
 Therefore each level of the pseudo spin projection will split to $(2I+1)$
nuclear spin projections with  the quantization axis $m=-I,\ldots
,I$.

This can couple to an electric field gradient (EFG) at the nuclear
position, expressed by the curvature of the crystal field
potential. The potential describing this coupling is written
\cite{Abr89}
\begin{equation}
V_Q \; = \;
\frac{-eQ}{I(2I-1)}[V_{11}I_{1}^{2}+V_{22}I_{2}^{2}+V_{33}I_{3}^{2}].
\label{eq8}
\end{equation}
The bases used here $(e_1,e_2,e_3)$ are the principal axes of the
tensor $V_{ij}$ which describes the electric field gradient, and
$e$ is the electron charge. According to the Laplace equation the
potential obey $V_{11}+V_{22}+V_{33}=0$. If we define the
asymmetry parameter $\eta = \frac{V_{22}-V_{33}}{V_{11}}$, the
quadrupole potential can be expressed as:
\begin{equation}
V_Q \; = \;\epsilon_q[3I_{1}^{2}+\eta (I^{2}_{2}-I_{3}^{2})-I^{2}]
\label{eq9}
\end{equation}
where we denote by $\epsilon_q=\frac{-eQ V_{11}}{4I(2I-1)}$ the
quadrupole coupling constant.

Therefore we can write the quadrupole potential in terms of the
reduced two-state coordinate:
\bea H_Q=\left[
V_Q^L(\frac{1+\sigma_z}{2})+V_Q^R(\frac{1-\sigma_z}{2})\right],
\label{eq10}
\eea
where $V_Q^{R(L)}$ is defined in Eq.~(\ref{eq8}) for the
particles in right (left) well \cite{Dor03}. We can go to basis
$|\psi_{\pm }(I,m_{\pm})\ket=|\psi_{\pm
}\ket\otimes|I,m_{\pm}\ket$ which have defined as
following\cite{Wue04}
\bea H_{\pm}|\psi_{\pm }(I,m_{\pm})\ket=E_{\pm,m_{\pm}} |\psi_{\pm
}(I,m_{\pm})\ket
\label{eq11}
\eea
where $H_{\pm}=H^D_{TLS}+(\frac{V_Q^L+V_Q^R}{2})\pm
w(\frac{V_Q^L-V_Q^R}{2})$ and therefore $E_{\pm,m_{\pm}}=\pm
\frac{E}{2}+\epsilon_{m_{\pm}}$; the corresponding eigen-states
satisfy:
\bea \bra I,m'_{\pm}|I,m_{\pm}\ket=\delta_{m'_{\pm},m_{\pm}}
\label{eq12} \eea
and since $H_+$ and $H_-$ do not commute,  their
eigen-states are not generally orthogonal:
\bea \bra I,m'_{\pm}|I,m_{\mp}\ket=\chi_{m'_{\pm},m_{\mp}}
\label{eq13}
\eea
where  these overlaps are dependent on the angle $\theta$. (here
$\theta$ is the angle between two axis of the Nuclear quadrupole
in each wells\cite{Akb06}: $\widehat{e_1^R,e_1^L }$)

\section{Heat Conductivity}
The dominant effect of uniform strain  field (describing the
interaction of the TLS with a phonon field)  is on the
energy of the tunneling state by changing the asymmetry energy.
The changes in the barrier height can usually be
ignored\cite{And-86}. Any external perturbation is therefore
diagonal in the local representation $(|L\ket,|R\ket)$ which when
transformed into the diagonal representation ( $|\psi_{+}\ket,
|\psi_{-}\ket$) has the form
\bea H_{int}= (\frac{\Delta _{0}}{E}\sigma _{x} + \frac{\Delta
}{E}\sigma _{z})\gamma e\cos(\omega t)
=H'_{int}\cos(\omega t)
\label{eq14}
\eea
in the presence of  a strain field $\xi=\xi_0\cos(\omega t)$,
where $\xi_0$  and $\omega$ are  the  amplitude and the frequency
of the strain field respectively.
 The strain is given by $e=\xi_0k_{\alpha}$,
  and the parameter $\gamma$, defined
as $\frac{1}{2}\frac{\partial\Delta}{\partial e}$, is equivalent
to
 elastic dipole moment. Where  $k_{\alpha}$ is the phonon
 wave-vector with polarization $\alpha$. Here the
tensorial nature of $e$ has been ignored and $\gamma e$ is written
as an average over orientations. Therefore we can easily
show\cite{phi87} that
$e=(\frac{\hbar}{2\rho\omega})^{\frac{1}{2}}k_{\alpha}$, where
 $\rho$ is the bulk density
and $\hbar$ is the Planck constant.

Using Fermi Golden Rule, one can obtain the contribution of a
phonon with wave vector $k_\alpha$ and polarization $\alpha$ to
the generalized  TLS transition probability due to phonon emission
and absorption, respectively:
\bea \Gamma^{em_{\alpha}}_{m'_+\longrightarrow
m_-}&=&\frac{2\pi}{\hbar}\mid
 \bra\psi_+(I,m'_+)\mid H'_{int}\mid\psi_-(I,m_-)\ket
\mid^{2}
\no&\times&
n_{E_{+,m'_+}}\delta(E_{+,m'_+}-E_{-,m_-}-\hbar\omega_{\alpha})\nonumber
\label{eq15}
\eea
and
\bea \Gamma^{ab_{\alpha}}_{m_-\longrightarrow
m'_+}&=&\frac{2\pi}{\hbar}\mid
 \bra\psi_-(I,m_-)\mid H'_{int}\mid\psi_+(I,m'_+)\ket
\mid^{2}
\no&\times&
n_{E_{-,m_-}}\delta(E_{+,m'_+}-E_{-,m_-}-\hbar\omega_{\alpha})\nonumber
\label{eq16}
\eea
where $n_{E_{\pm,m_\pm}}=e^{-\beta E_{\pm,m_\pm}}/Z$ is the
Boltzmann weight, $Z=\sum_{\pm,m_{\pm}} e^{-\beta E_{\pm,m_\pm}}
$, $\beta = \frac{1}{ K_BT}$, $K_B$ is the Boltzmann constant and
$T$ is temperature. It must be noted here that the transition between
the same TLS levels are zero:
\bea \bra\psi_-(I,m_{\pm})\mid
H'_{int}\mid\psi_+(I,m'_{\pm})\ket=0\longrightarrow\Gamma_{m_{\pm}\longrightarrow
m'_{\pm}}=0.\nonumber
\label{eq16b}\eea
Therefore the  phonon relaxation time can be found  by summing
$\Gamma^{ab_{\alpha}}_{m_-\longrightarrow m_+}-
\Gamma^{em_{\alpha}}_{m'_+\longrightarrow m'_-}$ over all
spin  states:
\bea \tau^{-1}_{\alpha}
=\sum_{m_-,m'_+}& &
\frac{2\pi\gamma_{\alpha}^{2}\omega_{\alpha}}{\rho v_{\alpha}^{2}}
u^2|\chi_{m'_+,m_-}|^2t_{m_-,m'_+}(E)
\no&&\times \delta[E-( \epsilon_{m_-}-
\epsilon_{m'_+}+\hbar\omega_{\alpha})],
\label{eq17}
\eea
where $v_{\alpha}$ is the sound velocity. Denoting
\bea t_{m_-,m'_+}(E)&=& n_{E_{-,m_-}}-n_{E_{+,m'_+}}
\\\nonumber
&=&
 \frac{e^{\beta E }e^{ -\beta \epsilon_{m_-}}-e^{-\beta
 \epsilon_{m'_+}}}{e^{\beta E}\sum_{m_-}e^{-\beta \epsilon_{m_-}}
+ \sum_{m'_+}e^{-\beta \epsilon_{m'_+}}}.
\label{eq18}
\eea
and after some calculations and averaging over TLS parameters
(using Eq.~\ref{eq3}), it can be easily shown that
\bea \tau^{-1}_{\alpha}&=&
\frac{P_0\pi\gamma_{\alpha}^{2}\omega_{\alpha}}{\rho
v_{\alpha}^{2}}
\\\nonumber&\times& \sum_{m_-,m'_+} |\chi_{m'_+,m_-}|^2 t_{m_-,m'_+}(
\epsilon_{m_-}- \epsilon_{m'_+}+\hbar\omega_{\alpha}).
\label{eq19}
 \eea
Neglecting the phase difference between the nuclear moments in
the two wells  and  assumming that the EFG in both wells are the same
$(\chi_{m'_+,m_-}=\delta_{m'_+,m_-}\Rightarrow\epsilon_{m_-}=
\epsilon_{m'_+})$,  the famous result of the
standard TLS model can be found:
\bea \tau^{-1}_{\alpha}&=&
\frac{P_0\pi\gamma_{\alpha}^{2}\omega_{\alpha}}{\rho
v_{\alpha}^{2}}
 \tanh(\beta
\hbar\omega_{\alpha}).
\label{eq20}
\eea
%

%
\begin{figure}[t]
\vspace{.25cm} \centering \scalebox{0.34}[0.34]{
\includegraphics*{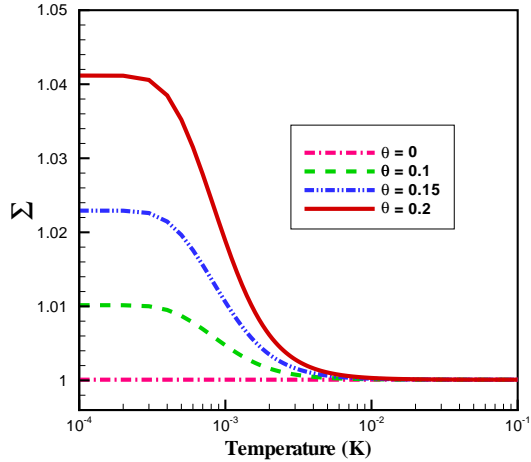}
}
\caption{ Thermal variation of the ratio parameter  $\Sigma$, for
$I=1$, $\epsilon_q=1 \mbox{mK}$ and the different angle between
the Nuclear Quadrupole moment in each TLS wells.
}
\label{fig1}
\end{figure}
%
The thermal conductivity $\kappa ( T )$
is evaluated on the assumption that heat is carried by
non-dispersive sound waves, therefore  one can write
\bea \kappa ( T )&=& \frac{1}{3} \sum_{\alpha}\int_{0}^{\infty}
l(\omega_{\alpha})C_{V}(\omega_{\alpha})g(\omega_{\alpha})v_{\alpha}d\omega_{\alpha},
\eea
where $l(\omega_{\alpha})=v_{\alpha}/\tau^{-1}_{\alpha} $ is the phonon mean free
path
 of angular frequency $\omega_{\alpha}$,
$g(\omega_{\alpha})=\frac{\omega_{\alpha}^{2}}{2\pi^{2}v_{\alpha}^{3}}$
is the phonon frequency distribution function, and
$C_{V}(\omega_{\alpha})$ is the heat capacity of phonon which
is given by
\bea C_{V}(\omega_{\alpha})=\frac{1}{(K_{B}T^{2})}
(\frac{\hbar\omega_{\alpha}}{2})^{2}\csch\!^{2}
(\frac{\beta\hbar\omega_{\alpha}}{2}).
\label{eq21}
\eea
By defining $x=\frac{\beta\hbar\omega_{\alpha}}{2}$
and using the above equations
the heat conductivity can be obtained,
\bea \kappa ( T )=  \Sigma(T)\times \kappa_{TLS}(T)
\label{eq22}
\eea
where 
$\kappa_{TLS}(T)=\sum_{\alpha}\frac{\rho
v_{\alpha}}{6\pi\hbar^2P_0\gamma_{\alpha}^2}K_B^3T^2$
 is the standard TLS (STLS)
heat conductivity\cite{black78}, and the coefficient $\Sigma(T)$ is defined by
\begin{widetext}
\
\bea
 \Sigma(T)= \frac{4}{\pi^2}
 \int_{0}^{\infty}\frac{x^3\csch^2(x)dx}{\sum_{m'_+,m_-}
|\chi_{m'_+,m_-}|^2 t_{m_-,m'_+}( \epsilon_{m_-}-
\epsilon_{m'_+}+2x/\beta)}, \no
\label{eq23}
 \eea
\end{widetext}

As  the exact
behavior of the Heat Conductivity can not be found  analytically, we are
trying to solve Eq.~\ref{eq23} numerically. 
Assuming that  $I=1$ and $\epsilon_q=1
\mbox{ mK}$ as suggested by echo experiments, we observed the behavior of parameter 
$\Sigma(T)$ in term of temperature. The results are peresented on Fig.~(\ref{fig1}) 
for different
values of 
quadrupole angle ($\theta$) and by averaging over the  $\eta$ parameter.

It can be seen that
in high temperature
regimes $(\epsilon_q\ll K_BT)$, this ratio $(\Sigma)$ goes to one. 
As it is predictable
where the nuclear part effect can be neglected and heat conductivity 
behaves as the Standard TLS model. 
Decreasing
 temperature  this ratio grows and will be saturated at very
low temperatures.

In agreement with expectation, at zero quadrupole angle the heat conductivity is 
the same as the result found from the standard TLS model
(please see Eq.\ref{eq19} and the statements before that).
At low temperature
regime, increasing the quadrupole angle with small value cause 
the heat conductivity 
 saturated value to  be larger than what is found form
the standard TLS model up to one percent;

The same behavior can be found  for $I=3/2 $ and $I=2$. Also it can be shown that  by
changing  $\epsilon_q$, the growing regime shows a
dependency on the quadrupole energy value; it means that by
increasing the magnitude of $\epsilon_q$, the growing regime will
be shifted to higher temperatures. It shows that there is a cross over
between two different regimes in the temperature  around the
quadrupole energy value.

\section{Summary and Conclusion}

In this paper we have studied the thermal properties of heat
conductivity  of cold glasses taking into account the quadrupole
effects. To describe the interaction of a TLS with nuclear
quadrupole, we have used a generalization of the standard TLS
Hamiltonian with nuclear spin. The nuclear quadrupole of these
systems leads to a splitting of the nuclear spin levels, which is
different for the ground state and the excited state. The presence
of this multi-level structure causes to increase the heat
conductivity magnitude compared to that of simple two-level
systems at lower temperature regimes ($K_BT\sim \epsilon_q$).

It is shown that the heat conductivity has a  cross over between
two different regimes  at the temperature which is close to the
nuclear quadrupole energy, $K_BT\sim \epsilon_q$. 
It can be observed that
there are three different regimes versus temperature.

 The first regime deals with high temperature regime: $K_BT\gg \epsilon_q$
 where the known
standard tunneling model results have been found. At this regime 
the effect of quadrupole energy can be
neglected  in comparison with the TLS
energy scale, therefore the nuclear spin splitting is not observable.

The second regime  demonstrates the temperature around nuclear
quadrupole energy: $K_BT\sim \epsilon_q$. In this regime,
 by decreasing the temperature the heat conductivity increases.
This is the cross over regime between the standard TLS behavior and
the low temperature regime where the nuclear quadrupole effects become
important.
In this area the nuclear quadrupole  energy  levels play an important role 
in the thermal behavior of the system. Decreasing  temprature the nuclear quadrupole energy
is comparable to thermal fluctuations.

In general these sub-energy level are not the same in both wells of  the TLS.
Thus  their eigen-states are not orthogonal and have the overlap with each other
 ($|\chi_{m'_+,m_-}|^2\neq \delta{m'_+,m_-}$ and $\epsilon_{m_+}\neq\epsilon_{m_-}$).
This effect causes the mean free path of phonons to increase
therefore the thermal conductivity has larger value 
in comparison with the simple two level system at low temprature regime.
This means that where  $K_BT\sim \epsilon_q$, $\Sigma=1+\epsilon (\theta)$ and 
heat conductivity exponent, $p$, is less than two instead of the
$p=2$ which has been found for standard TLS model.

Finally for the third regime the heat conductivity will be saturated at
$K_BT\ll \epsilon_q$.

To obtain a theorytical expression for this effect, 
one can write $\chi_{m'_+,m_-}=\delta_{m'_+,m_-}+\varsigma_{m'_+,m_-}$ and 
$\epsilon_{m_+}=
\epsilon^o_{m_+}+\gamma_{m_+}$ where $\epsilon^o_{m_+}=\epsilon_{m_-}$,
and move to the special limitting case 
where $\varsigma_{m_+,m_-}\ll\beta\gamma_{m_+} $, and $\gamma_{m_+}\ll \epsilon^o_{m_+}$.
By a little manipulation it can  be easily shown that 
\
\bea
\epsilon (\theta)= C\times (\delta \beta \gamma)^2
\label{eq24}
 \eea
where 
$C=\frac{1}{256} \left[48 \pi ^2+\pi ^4-384 \zeta (3)\right]\approx 0.428$ 
is a numerical constant;
 $(\delta \beta \gamma)^2=\bra \beta^2\gamma^2 \ket-\bra \beta\gamma \ket^2$,  
$\bra \beta \gamma\ket=\sum_{m_+}e^{-\beta
\epsilon^o_{m_+}}\beta\gamma_{m_+}/\sum_{m_+}e^{-\beta
\epsilon^o_{m_+}}$, and 
$\bra\beta^2 \gamma^2\ket=\sum_{m_+}e^{-\beta
\epsilon^o_{m_+}}\beta^2\gamma_{m_+}^2/\sum_{m_+}e^{-\beta
\epsilon^o_{m_+}}$.
It shows clearly that by increasing the quadrupole angle the difference of
 sub-energy in both wells increases which causes 
the $\Sigma(T)$ value to increase, in agreement with numerical results.

In conclusion we believe that nuclear quadrupoles  play an
important role in the nature of glasses at low temperatures. In
this respect  for solving the problem of cold glasses, it is
useful to find the effect of nuclear spin on the other response
functions. As far as we know there is no experimental result for
the heat conductivity\cite{Pohl02} in the case $K_BT\sim
\epsilon_q$. Therefore it might be a good suggestion for future
experiments to approach lower temperatures or use the glasses with
larger quadrupole energy.


\begin{acknowledgments}
I would like to express my deep gratitude to  A. Langari  for
stimulating discussions and useful comments. I am also grateful to
A. W\"{u}rger and M. Aliee for the fruitful discussions.
\end{acknowledgments}


\end{document}